\documentclass[aps,amsmath,twocolumn,amssymb,titlepage,10pt]{revtex4-1}
\usepackage[T1]{fontenc}
\usepackage[utf8]{inputenc}
\usepackage{amsmath}
\usepackage{braket}
\usepackage{amsfonts}
\usepackage{graphicx}
\usepackage{hyperref}

\usepackage{amssymb}
\usepackage{amsmath}

\begin{document}
\title{Spin vs. bond correlations along dangling edges of quantum critical magnets}
\author{Lukas Weber}
\author{Stefan Wessel}
\affiliation{Institute for Theoretical Solid State Physics, JARA-FIT, and JARA-HPC, RWTH Aachen University, 52056 Aachen, Germany}
\date{\today}
\begin{abstract}
Dangling edge spins of two-dimensional quantum critical antiferromagnets display strongly enhanced spin correlations with scaling dimensions that fall outside of
the classical theory of surface critical phenomena. We provide large-scale quantum Monte Carlo results for both spin and bond correlation functions for the case of the columnar dimer model in particular. Unlike the spin correlations, we find the bond correlations to differ starkly between the spin-1/2 and spin-1 case. Furthermore, we compare the corresponding scaling dimensions to recent theoretical predictions. These predictions are, in part, supported by our numerical data, but cannot explain our findings completely. Our results thus put further constraints on completing the understanding of dangling edge correlations, as well as surface phenomena in strongly-correlated quantum systems in general.
\end{abstract}
\maketitle
The surface physics of quantum systems has become a central topic in condensed matter research in recent years. For example, topological phases of matter with robust edge states hold the promise for practical applications in quantum computing~\cite{Mourik2012,Albrecht2016,Karzig2017}.
Increasingly, topological properties of strongly correlated systems are also moving into the focus of attention~\cite{Broholm2020,Chen2012}. Indeed, even in the absence of topological edge states, strongly interacting systems can host nontrivial surface effects. Well-known examples are provided already by the plethora of surface critical phenomena in classical systems~\cite{Domb1986},
where nontrivial surface correlations accompany a bulk critical point. By the quantum-to-classical mapping, a quantum equivalent of these classical effects should exist, which also motivated questions about the interplay between surface critical phenomena and the edge states of topological insulators~\cite{Grover2012,Parker2018}.

Within the classical theory based on the O($N$) model, when the bulk is critical, different classes of surface criticality exist. These surface universality classes, dubbed \textit{ordinary}, \textit{special} and \textit{extraordinary}, differ in the critical exponents they display in the decay of their surface correlations. At the ordinary transition, surface and bulk develop order simultaneously, whereas at the extraordinary transition, surface order is already present. Finally, the special transition denotes a multicritical point at which the   ordinary and extraordinary transition lines meet. This picture is modified in lower dimensions, as the Mermin-Wagner theorem~\cite{Mermin1966} may forbid stand-alone surface order. In such a case, relevant for instance in the three-dimensional (3D) O($3$) model, the special and extraordinary classes are thought to be absent, giving way to a purely \textit{ordinary} surface phase diagram.

Recent numerical studies of two-dimensional quantum Heisenberg antiferromagnets (AFM) have, however, called the current understanding of quantum surface critical phenomena into question~\cite{Zhang2017,Ding2018,Weber2018,Weber2019,Zhu2020}. These magnets feature a bulk quantum critical point in the 3D  O(3) universality class, precisely implementing the purely ordinary case discussed above. While for certain edge-spin configurations, the classical ordinary predictions for the surface critical phenomena are indeed recovered, edge configurations with dangling spins however display strikingly enhanced edge correlations. This disagreement holds fairly generally across different models and, surprisingly, both for spin-1/2 and spin-1 degrees of freedom~\cite{Weber2019,Zhu2020}. 
The observation that also spin-1 systems exhibits such enhanced edge spin correlations
is inconsistent with an earlier rationalization of the enhanced spin correlations by the presence of a topological $\theta$ term in the effective action of the spin-1/2 edge spin system~\cite{Zhang2017}, and called for further theoretical investigations. 

More recently, a couple of field theoretical scenarios have indeed been proposed to interpret the numerical observations. For the $S=1/2$ case, Jian~\textit{et al.}~\cite{Xu2020} suggest that the enhancement in the edge correlations stems from the strong correlations of the stand-alone dangling-edge-spin chain. By coupling an SU(2)$_1$ conformal field theory, describing the isolated chain, to the surface correlations of a classical critical bulk, they obtain a renormalization group flow  with a metastable fixed point. This fixed point corresponds to a direct, continuous critical point between Néel and valence-bond-solid (VBS) order. The parameter regime on the Néel-side of the critical point is affected by slow and nonmonotonic flow trajectories. In such a scenario, dangling edge  spin quantum magnets would be situated in a parameter region that will eventually flow to the Néel fixed point, but with finite-size properties that are governed by the metastable fixed point.

 Metlitski~\cite{Metlitski2020}  instead finds, based on an renormalization group analysis, that the classical understanding of the semi-infinite O($N$) model in three dimensions may be incomplete: Among different possible scenarios, within a restricted range of $N$, the special and extraordinary classes may exist even in the absence of pure surface order. In particular, this extraordinary class may display power-law surface correlations over a  range of $N$, including $N=3$, hereby possibly explaining the nonordinary correlations observed among the dangling edge spins. Apart from the spin correlations, both proposals also make distinct predictions for the bond correlations along the edge, to be detailed further below, which are amenable to a numerical assessment.

\begin{figure}
	\includegraphics{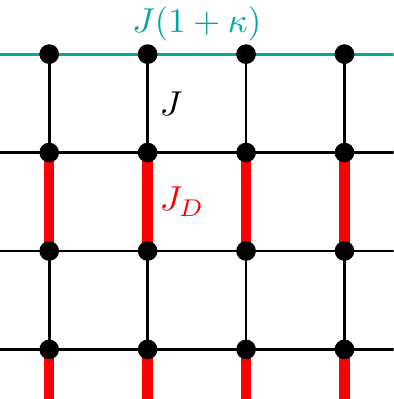}
	\caption{The columnar-dimer model with dimer bonds $J_D$ (thick, red) and interdimer bonds $J$ (thin, black). At the edge, the spins are dangling in the sense that they do not belong to a dimer. The coupling between the edge spins is further modified by a factor ($1+\kappa$).}
	\label{fig:lattice}
\end{figure}

Here, we provide numerical results for the bond correlations and compare them to the above theoretical predictions. As a particular realization of a dimerized Heisenberg AFM, we consider  the columnar dimer lattice with dangling edge spins, in which the dimer bonds form columns on a square lattice, and the edge spins remain unpaired (cf.~Fig.~\ref{fig:lattice}).
The Hamiltonian of the model in terms of spin-$S$ degrees of freedom reads
\begin{align}
	H &= J_D \sum_{x,y} \mathbf{S}_{x,y}\cdot \mathbf{S}^\prime_{x,y}\\
	&+ J \sum_{x,y} \mathbf{S}_{x,y-1} \cdot \mathbf{S}^\prime_{x,y} + \mathbf{S}_{x,y} \cdot \mathbf{S}_{x+1,y} + \mathbf{S}^\prime_{x,y} \cdot \mathbf{S}^\prime_{x+1,y}\nonumber\\
	&+ J (1+\kappa) \sum_x \mathbf{S}_{x,0} \cdot \mathbf{S}_{x+1,0}\nonumber,
\end{align}
where $J, J_D > 0$ and $\mathbf{S}_{x,y}$ and $\mathbf{S}^\prime_{x,y}$ denote the two spins forming a dimer at unit cell position $(x,y)$. In the sums, $y$ starts from $1$, so that the edge is formed by the spins $\mathbf{S}_{x,0}$. In the bulk, this model features a Néel-ordered phase (for $J\approx J_D$) and a quantum-disordered singlet-product phase ($J \ll J_D$). These two phases are separated by a quantum critical point of the 3D O(3) universality class, situated at $J/J_D = 0.523\,37(3)$~\cite{Matsumoto2001,Wenzel2008} for $S=1/2$, and $J/J_D = 0.189\,20(2)$~\cite{Weber2019} for $S=1$, respectively. In the simulations reported below, the couplings $J$ and $J_D$ were fixed to these ratios in order to to tune the bulk to the quantum critical point. Furthermore, the surface spin coupling contains a factor $(1+\kappa)$ that allows tuning the ratio between the surface and the bulk coupling.

The surface spin and bond correlations along the dangling edge spins that we examine below are defined as
\begin{align}
	C^n_\parallel(r) &= \braket{\mathbf{S}_{0,0}\cdot \mathbf{S}_{r,0}},\\
	C^v_\parallel(r) &= \braket{\left(\mathbf{S}_{0,0}\cdot \mathbf{S}_{1,0}\right)\left(\mathbf{S}_{r,0}\cdot \mathbf{S}_{r+1,0}\right)} - \braket{\mathbf{S}_{0,0}\cdot \mathbf{S}_{1,0}}^2.
\end{align}
When the bulk is quantum critical, these surface correlation functions are expected to admit power-law scaling, 
\begin{equation}
	|C^{n/v}_\parallel(r)| \sim r^{-2\Delta_{n/v}}
	\label{eq:powerlaw}
\end{equation}
at large distances $r$, as is the case for their bulk equivalents. The exponents of the power law decay are given by the scaling dimensions $\Delta_n$ and $\Delta_v$ of the Néel and VBS order parameters, respectively.
The Néel scaling dimension  relates via $\Delta_n=(d+z-2+\eta_\parallel)/2$ to the  exponent $\eta_\parallel$ determined in previous studies for $\kappa=0$~\cite{Ding2018,Weber2018,Weber2019}. While simulations for $\kappa \ne 0$  were performed previously~\cite{Weber2019}, a detailed scaling analysis of the critical exponents was not carried out. In this work, we  take a closer look at a range of values for $\kappa$ as well as larger system sizes up to $160\times160$ unit cells. We then extract and compare the corresponding exponents $\Delta_n$ and $\Delta_v$ to the field-theoretical predictions.

For the VBS scaling dimension $\Delta_v$, the field-theoretical scenarios mentioned earlier make distinct predictions.
On the one hand, at the Néel-VBS critical point of Ref.~\cite{Xu2020}, $\Delta_v$ and $\Delta_n$ show deviations from their SU(2)$_1$ values,
\begin{align}
    \Delta_n - \frac{1}{2} &= \epsilon_n,\\
    \label{eq:epsn}
    \Delta_v - \frac{1}{2} &= - 3 \epsilon_n,
\end{align}
to leading order in $\epsilon_n = 3/2 - \Delta_\Phi$. The scaling dimension $\Delta_\Phi$ of the classical surface correlations is set to the value for the \textit{ordinary} class $\Delta_\Phi^\text{ord}\approx 1.2$~\cite{Deng2005}. Upon eliminating $\epsilon_n$, one obtains a scaling relation connecting $\Delta_n$ and $\Delta_v$,
\begin{equation}
    \Delta_v = -3 \Delta_n +2 .
    \label{eq:scaling_relation}
\end{equation}
On the other hand, Ref.~\cite{Metlitski2020} argues that at the extraordinary-power-law transition, the nonperturbative skyrmion contributions to the partition function could be numerically small. This would explain the similarity of $\Delta_n$ between $S=1/2$ and $S=1$, and, in addition, suppresses perturbative corrections on $\Delta_v$, i.e., a value of 
\begin{equation}
    \Delta_v = 2,
    \label{eq:deltav2}
\end{equation}
up to nonperturbative effects.
%
To test these theoretical predictions, we performed quantum Monte Carlo
simulations using the stochastic series expansion (SSE) with directed-loop updates~\cite{Sandvik2002,Syljuasen2003}. In order to probe ground state properties, we scaled the temperature  as $T=J_\text{min}/2L$, and performed finite-size scaling in the linear system size $L$ and inverse temperature $1/T$, respecting a dynamical critical exponent of  $z=1$. Here, $J_\text{min}$ is the smallest coupling of interest, which in our case equals $J_\text{min} = J(1+\kappa_\text{min})= 0.3 J$. Taking this factor into account is necessary to obtain the correct ground state behaviour when $\kappa$ approaches $-1$, i.e., the regime of weak intraedge coupling. 

\begin{figure}
	\includegraphics{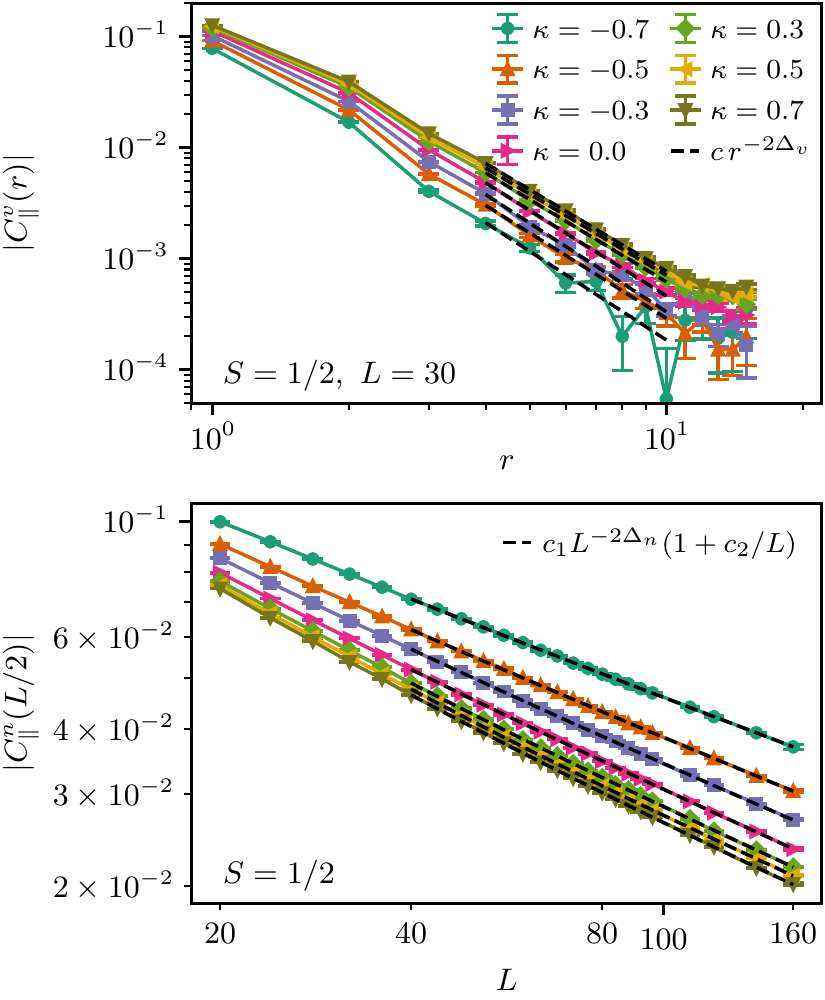}
	\caption{Boundary bond correlations $C^v_\parallel(r)$ at fixed system size (top panel) and spin correlations $C^n_\parallel(L/2)$ as a function of $L$ (bottom panel), both for different values of the edge coupling $\kappa$. The dashed lines show the fits that were used to extract the scaling dimensions as detailed in the main text.}
	\label{fig:kappa_corr}
\end{figure}

The spin correlations $C^n_\parallel$ are easily accessible within the SSE in terms of their diagonal part in our computational $S^z$ basis, given the SU(2) symmetry of $H$. The bond correlations $C^v_\parallel$ can be measured using the fact that the bond operator itself is contained in the Hamiltonian, allowing the use of an operator-count estimator~\cite{Sandvik1992}.
In the case of $C^n_\parallel$, we calculated the maximum-distance correlations $C^n_\parallel(L/2)$ for different system sizes. In the case of $C^v_\parallel$, however, the estimator is severely affected by poor statistics, restricting us to examine the full fixed-$L$ $C^v_\parallel(r)$ correlations at several smaller system sizes.
Figure~\ref{fig:kappa_corr} shows the SSE results for both correlation functions for a range of $\kappa$ values. At first glance, we find that while the spin correlations are enhanced upon lowering $\kappa$, the  competing bond correlations decrease correspondingly. The slope in the log-log plot, however, does not change drastically in the range of $\kappa$ studied for both correlation functions.

In the next step, we extract the scaling dimensions $\Delta_n$ and $\Delta_v$ using power-law fits. As the scaling forms given in Eq.~\eqref{eq:powerlaw} hold only in the thermodynamic limit, for the fits of the finite-size data, we need to take finite-size scaling corrections into account. For $C^n_\parallel$, we have access to a large range of system sizes and minimize the finite-size effects by only taking systems with $L\ge 40$ into account. Moreover, we include an effective correction to scaling into our fit form
\begin{equation}
	C^{n,\text{fit}}_\parallel(L/2) = c_1 \,(-1)^{L/2}\,L^{-2 \Delta_n} \left(1+\frac{c_2}{L}\right),
\end{equation}
where the factor $(-1)^{L/2}$ accounts for the sublattice structure of the AFM correlations, and
the factors $c_1$, $c_2$ are nonuniversal fit parameters. For the bond correlations, the situation is less favorable. At fixed system size, the decay of $C_\parallel^v(r)$ is eventually limited by the periodic boundary conditions parallel to the edge. We can therefore expect to observe the asymptotic power-law decay only within an intermediate regime $1 \ll r \ll L/2$. In practice, we choose the range $4 \le r \le L/3$, in which we fit to a simple power law 
\begin{equation}
	C^{v,\text{fit}}_\parallel(r) = c\,(-1)^{r}\,r^{-2 \Delta_v},
\end{equation}
with $c$ a  nonuniversal fit parameter.
The resulting exponents $\Delta_v$ are still affected by finite-size effects, the magnitude of which we can estimate by performing the same analysis for several system sizes  $L=30,40$, and $60$. 

\begin{figure}
	\includegraphics{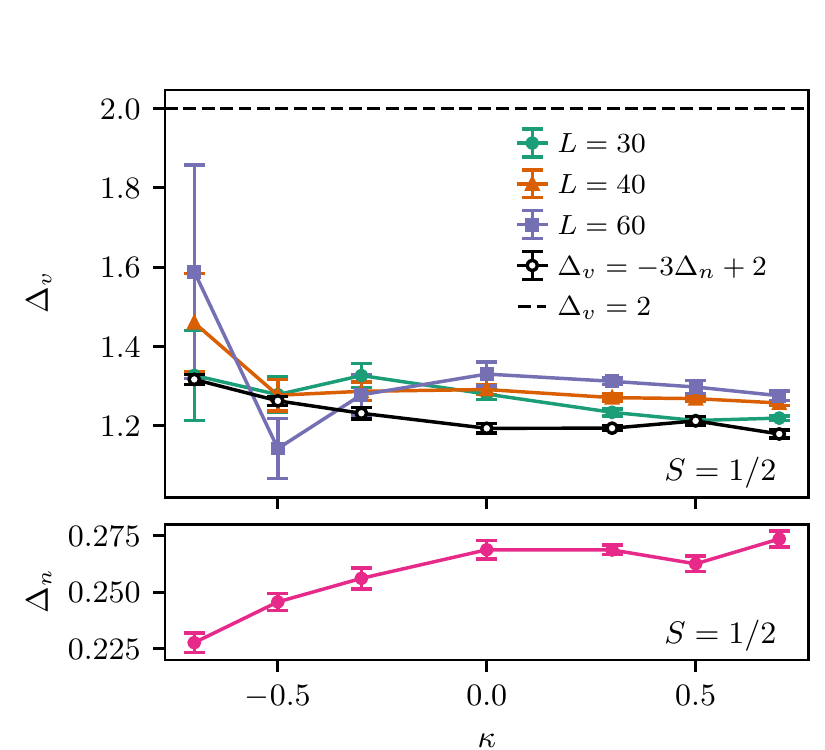}
	\caption{$S=1/2$ fit results for the exponents $\Delta_v$ (top panel) and $\Delta_n$ (bottom panel), as extracted from the data in Fig.~\ref{fig:kappa_corr} as a function of $\kappa$. The error bars show the statistical error of the fits, while additional finite-size effects for $\Delta_v$ are observed in the indicated $L$-dependence.  For comparison, an estimate based on the scaling relation \eqref{eq:scaling_relation} in the main text~\cite{Xu2020} and the estimate $\Delta_v = 2$ from Ref.~\cite{Metlitski2020} are shown.}
	\label{fig:deltas}
\end{figure}

In Figure~\ref{fig:deltas}, the resulting exponents are shown, along with the field theoretical estimates according to Ref.~\cite{Xu2020} and Ref.~\cite{Metlitski2020}.
On the one hand, while the obtained values of $\Delta_v$ are not converged to the thermodynamic limit, our numerical values fall well below the prediction $\Delta_v=2$ given in Ref.~\cite{Metlitski2020}. On the other hand, the scaling relation from Eq.~\eqref{eq:scaling_relation}, connecting $\Delta_v$ and $\Delta_n$, roughly agrees with the obtained exponents. Furthermore, we find that $\Delta_n$ decreases for $\kappa<0$. Since lowering $\kappa$ corresponds to an effective (relative) increase of the edge-bulk coupling, these results are in line with the flow diagram proposed in Ref.~\cite{Xu2020}.

\begin{figure}
	\includegraphics{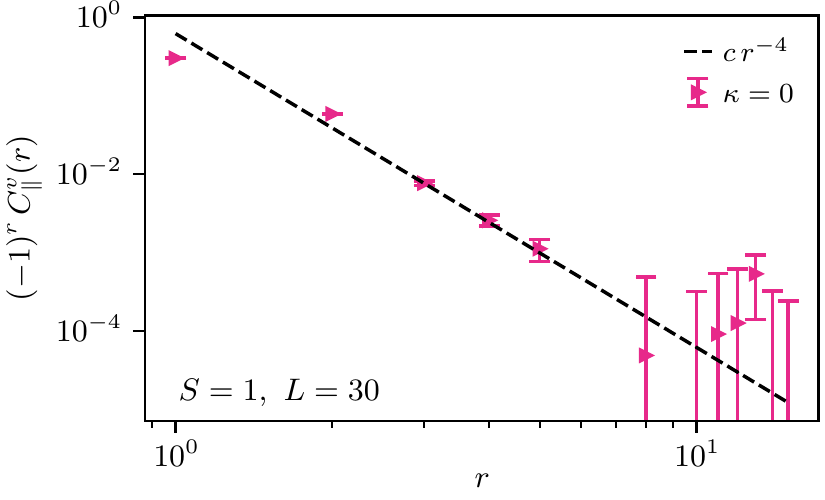}
	\caption{$S=1$ bond correlations $C^v_\parallel(r)$ at $L=30$. Due to the strong decay, only few values can be distinguished from zero. As a rough estimate, the decay is compared to $r^{-4}$ (dashed line) which would correspond to $\Delta_v=2$.}
	\label{fig:spin1}
\end{figure}

For $S=1$, the bond correlations display a much stronger decay, rendering the statistical uncertainty more severe. We therefore concentrate on $\kappa=0$ and set $T=J_\text{min}/2L = J/2L$ (cf.~Fig.~\ref{fig:spin1}). We can resolve finite correlations only at short distances, which alone is not enough to extract the true long-range behavior. However, the data is compatible with $\Delta_v=2$, which corresponds to a $r^{-4}$ decay. While the match is plausible, other forms of decay cannot be excluded.

%
%
In summary, 
we computed the surface spin and bond correlations for the dangling columnar dimer model and compared their corresponding  scaling dimension to the two distinct field-theory predictions from Refs.~\cite{Xu2020,Metlitski2020}. While for $S=1/2$ we find indications for a scaling relation relating the VBS scaling dimension $\Delta_v$ with the Néel scaling dimension $\Delta_n$, for $S=1$ our results are consistent with $\Delta_v=2$, obtaining no anomalous dimension. Thus, unlike the spin correlations, which were previously found to be similar between $S=1/2$ and $S=1$, the bond correlations strongly depend on the values of the quantum spin $S$.

In the context of Ref.~\cite{Metlitski2020}, these results suggest that for the $S=1/2$ case, nonperturbative effects do play a role. It is yet unclear, however, why these effects influence $\Delta_v$, but seemingly do not affect $\Delta_n$. 
For $S=1/2$, the bond correlations may indeed be governed by the physics of the fixed point found in Ref.~\cite{Xu2020}, explaining the reasonable agreement of the scaling relation in Eq.~\eqref{eq:scaling_relation} with our data. However, at the root of this relation is the assumption that there is an underlying degree of freedom with scaling dimension $\Delta_\Phi = 3/2-\epsilon_n$ that couples to the surface field. In Ref.~\cite{Xu2020}, this scaling dimension was identified with the classical ordinary surface scaling dimension $\Delta_\Phi^\text{ord}\approx 1.2$~\cite{Deng2005}, see also the discussion in Ref.~\cite{Metlitski2020}.  This identification can be motivated by the microscopic coupling of the dangling-edge-spin chain coupled to a nondanging, ordinary quantum magnet. However, such a positive value for $\epsilon_n$ would lead to an enhancement of the bond correlations, compared to the spin correlations (cf.~Eq.~\eqref{eq:epsn}). We instead observe the opposite effect, which could be explained by a negative value of $\epsilon_n$, corresponding to $\Delta_\Phi^\prime\approx 1.75$. It is not obvious, how such a degree of freedom would appear in the considered model.
These issues show that there are still gaps in the current theoretical understanding of these phenomena. We hope that based on our findings, future treatments can yield a unified understanding of the surface criticality of dangling edge quantum magnets. \

We thank C.-M. Jian and C. Xu, M. A. Metlitski, as well as F. P. Toldin for insightful discussions, and the suggestion to examine the bond correlations (M. A. Metlitski) and to probe the scaling relation (C. Xu). Furthermore, we acknowledge support by the DFG through Grant No. WE/3649/4-2 of the FOR 1807 and through RTG 1995, and thank the IT Center at RWTH Aachen University and the JSC Jülich for access to computing time through JARA-HPC.

\bibliography{paper}

\end{document}